\documentstyle[12pt]{article}
\begin {document}
\baselineskip 2.25pc

\begin{center}
{\Large Levinson theorem in two dimensions}
\vskip 2pc
 Qiong-gui Lin\\
        China Center of Advanced Science and Technology (World
	Laboratory),\\
        P.O.Box 8730, Beijing 100080, P.R.China\\
        and\\
        Department of Physics, Zhongshan University, Guangzhou
        510275, P.R.China \footnote{Mailing Address}

\end{center}
\vfill
\centerline{\bf Abstract}

A two-dimensional analogue of Levinson's theorem for nonrelativistic
quantum mechanics is established, which relates the phase shift at
threshold(zero momentum) for the $m$th partial wave to the total
number
of bound states with angular momentum $m\hbar(m=0,1,2,\ldots)$ in an
attractive central field.

\vfill

\leftline {PACS number(s): 34.10.+x, 34.90.+q, 03.65.-w}

\newpage

\section*{\large \S1. Introduction}

In 1949, a theorem in quantum mechanics was established by
Levinson[1],
which relates the phase shift at threshold(zero momentum) for the
$l$th partial wave, $\delta_l(0)$, to the number of bound states
with the same
azimuthal quantum number, $n_l$. This is one of the most interesting
and beautiful results in nonrelativistic quantum theory. The subject
was then studied by many authors, some are listed in the References%
[2-5]. The relativistic generalization of Levinson's theorem has also
been well established[6-8]. However, most of these authors deal with
the problem in ordinary three-dimensional space. To our knowledge
a two-dimensional version of Levinson's theorem was not presented
in the literature. The purpose of this work is to develop an analogue
of this theorem in two spatial dimensions. It relates the phase shift
at threshold for the $m$th partial wave, $\eta_m(0)$, to the total
number of bound states with angular momentum $m\hbar$, $n_m$(the
total number of bound states with angular momentum $-m\hbar$ is also
$n_m$):
$$\eta_m(0)=n_m\pi, \quad m=0,1,2,\ldots.$$
This is similar to but slightly simpler than the original one in
three dimensions. In three dimensions Levinson's theorem takes the
same
form with $m$ replaced by $l$. But when $l=0$ the relation must be
modified if there exists a zero-energy resonance(a half-bound state).
The mathematical origin is that the behaviour of the phase shifts
$\delta_l(k)$ near
$k=0$ may be different for $l=0$ and $l\ne0$. In two dimensions
no similar situation occurs. This paper is organized as follows. In
the next section we give a brief formulation of the partial-wave
method for nonrelativistic scattering in
two spatial dimensions. In \S3
we discuss the behaviour of the phase shifts near $k=0$ in some
detail.
In \S4 we establish the Levinson theorem using the Green function
method[2,3,5]. \S5 is devoted to the discussion of some aspects
of the theorem.

\section*{\large \S2. Partial-wave method in two dimensions}

A particle with mass $\mu$ and energy $E$ moving in an external
field $V({\bf r})$ satisfies the stationary Schr\"odinger
equation
\begin{equation}
H\psi=-{\hbar^2\over2\mu}\nabla^2\psi+V({\bf r})\psi=E\psi.
\end{equation}
We use the polar coordinates $(r,\theta)$ as well as the rectangular
coordinates $(x, y)$ in two spatial dimensions. For scattering
problems $E>0$(we assume that $V\to 0$ more rapidly than $r^{-2}$
when $r\to\infty$).
The incident wave may be chosen as
\begin{equation} \psi_i=e^{ikx} \end{equation}
which solves Eq.(1) when $r\to\infty$ provided that $k=\sqrt{2\mu E/
\hbar^2}$. The scattered wave should have the asymptotic form
\begin{equation}
\psi_s\stackrel{r\to\infty}{\longrightarrow}\sqrt{i\over r}
f(\theta)e^{ikr} \end{equation}       %3
where the factor $\sqrt i=e^{i\pi/4}$ is introduced for later
convenience. This also solves (1) when $r\to \infty$. It is easy to
show that the differential cross section $\sigma(\theta)$(in two
spatial dimensions the cross section may be more appropriately called
cross width) is given in terms of the scattering amplitude
$f(\theta)$ by
\begin{equation} \sigma(\theta)=|f(\theta)|^2.\end{equation}
The outgoing wave comprising (2) and (3) and thus takes the following
form at infinity.
\begin{equation} \psi\stackrel{r\to\infty}{\longrightarrow}e^{ikx}+
\sqrt{i\over r} f(\theta)e^{ikr}. \end{equation}       %5
We are interested in central or spherically symmetric
(actually cylindrically
symmetric in two dimensions) potentials $V({\bf r})=V(r)$.
In this paper we deal only with central potentials. Then solutions
of Eq.(1) may be expanded as
\begin{equation}
\psi(r,\theta)=\sum_{m=-\infty}^{+\infty}a_mR_{|m|}(r)
e^{im\theta} \end{equation}
where $R_m(r)$ satisfies the radial equation
\begin{equation}
R''_m+{1\over r}R'_m+\left(k^2-{2\mu\over\hbar^2}V-{m^2\over r^2}
\right)R_m=0,\quad m=0,1,2,\ldots.\end{equation}          %7
If $V=0$, the regular solution of (7) is the Bessel function and may
be taken as $R_m^{(0)}(r)=\sqrt kJ_m(kr)$ and hence has the asymptotic
form
\begin{equation} R_m^{(0)}(r)\stackrel{r\to\infty}{\longrightarrow}
\sqrt{2\over \pi r} \cos\left(kr-{m\pi\over 2}-{\pi\over 4}\right).
\end{equation}       %8
We have assumed that $V(r)\to0$ more rapidly than $r^{-2}$
when $r\to\infty$,
so the solution $R_m(r)$ approaches the form of a linear combination
of the
Bessel function and the Neumann function at large $r$ and may have
the asymptotic form
\begin{equation} R_m(r)\stackrel{r\to\infty}{\longrightarrow}
\sqrt{2\over \pi r} \cos\left[kr-{m\pi\over 2}-{\pi\over 4}
+\eta_m(k)\right].   \end{equation}       %9
Here $\eta_m(k)$ is the phase shift of the $m$th partial wave. It is a
function of $k$. As in three dimensions, all $\eta_m(k)$ are real in a
real potential. Substituting (9) into (6) gives one asymptotic form
for $\psi$, while substitution of the formula
\begin{equation} e^{ikx}=\sum_{m=-\infty}^{+\infty}i^{|m|}J_{|m|}(kr)
e^{im\theta} \end{equation}                                      %10
and the asymptotic form of the Bessel functions into (5) gives
another. Comparing these two asymptotic forms one finds an expression
for $f(\theta)$ in terms of the phase shifts:
\begin{equation}
f(\theta)=\sum_{m=-\infty}^{+\infty}\sqrt{2\over\pi k}e^{i\eta_{|m|}}
\sin\eta_{|m|}e^{im\theta}. \end{equation}                                      %11
The total cross section $\sigma_t$ turns out to be
\begin{equation}
\sigma_t=\int_0^{2\pi}d\theta\,\sigma(\theta)={4\over k}
\left(\sin^2\eta_0+2\sum_{m=1}^\infty\sin^2\eta_m\right).
\end{equation}
From the above relations one easily realizes that all information of
the scattering process is contained in the phase shifts $\eta_m(k)$.
The latter are determined by solving the radial equation (7) with
the boundary condition (9) and thus depend on the particular form
of $V(r)$. In an attractive field, the number of bound states with
given angular momentum $m\hbar$, denoted by $n_m$ above, also depends
on the particular form of $V(r)$. It will be shown that $n_m$ is
related to $\eta_m(k)$ at threshold. This is similar to Levinson's
theorem in three dimensions. In the next section we first discuss the
behaviour of $\eta_m(k)$ near $k=0$.

\section*{\large \S3. Phase shifts near threshold}

Assuming that $V(r)$ is less singular than $r^{-2}$ when $r\to0$,
then the regular solution of the radial equation (7) may have the
following power dependence on $r$ near $r=0$.
\begin{equation}
f_m(r,k)\stackrel{r\to0}{\longrightarrow}{r^m\over 2^m m!},
\quad m=0,1,2,\ldots.\end{equation}
Here we denote the regular solution of (7) with the boundary condition
(13) by $f_m(r,k)$. Note that the equation (7) depends on $k$ only
through $k^2$, which is an integral function of $k$, and the boundary
condition (13) is independent of $k$. Thus a theorem of Poincar\'e
tells us that $f_m(r,k)$ is an integral function of $k$ for a fixed
$r$. On the other hand, the solution $R_m(r)$ with the boundary
condition (9), which is proportional to $f_m(r,k)$, need not be an
integral function of $k$. We denote a potential that satisfies
$V(r)=0$ when $r>a>0$ by $V_a(r)$. In such potentials the solution
of (7) when $r>a$ may take the form
\begin{equation}
R_m^+(r)=\sqrt k[\cos\eta_m J_m(kr)-\sin\eta_m N_m(kr)],\quad
m=0,1,2,\ldots \end{equation}                %14
where the superscript ``+'' indicates $r>a$. It is easy to verify
that $R_m^+(r)$ indeed satisfies the boundary condition (9). When
$r<a$ we have
\begin{equation} R_m^-(r)=A_m(k)f_m(r,k),\quad m=0,1,2,\ldots
\end{equation}     %15
where the superscript ``$-$'' indicates $r<a$. In general the
coefficient
$A_m$ depends on $k$, so that the two parts of $R_m(r)$ can be
connected smoothly at $r=a$. This  leads to
\begin{equation} \tan\eta_m={\rho J'_m(\rho)-\beta_m(\rho)J_m(\rho)
\over \rho N'_m(\rho)-\beta_m(\rho)N_m(\rho)}\end{equation}
where $\rho=ka$ and
\begin{equation} \beta_m(\rho)={af'_m(a,k)\over f_m(a,k)}
\end{equation}
where the prime indicates differentiation with respect to $r$. As
mentioned above,  $f_m(a,k)$ is an integral function of $k$, so is
$f'_m(a,k)$. Moreover, both of them are even functions of $k$ since
Eq.(7) depends only on $k^2$. Therefore when $k\to0$ or $\rho\to0$,
the leading term for $\beta_m(\rho)$ may have one of the following
forms
$$\beta_m(\rho)\to \alpha_m^+\rho^{2l_m^+},$$
$$\beta_m(\rho)\to \alpha_m^-\rho^{-2l_m^-},$$
$$\beta_m(\rho)\to \gamma_m+\alpha_m \rho^{2l_m}$$
where $\alpha_m^\pm$, $\alpha_m$, and $\gamma_m$ are
nonzero constants, while $l_m^\pm$ and $l_m$ are natural numbers.
Using these relations and the leading terms of
$J_m(\rho)$ and $N_m(\rho)$ for $\rho\to0$, the leading term in
$\tan\eta_m$
when $\rho\to0$ can be explicitly worked out. When $\gamma_m=-m$ some
care should be taken. However, careful analysis gives in any case
\begin{equation} \tan\eta_m\to b_m \rho^{2p_m}\quad{\rm or}\quad
{\pi\over 2\ln \rho}\quad(k\to 0) \end{equation}     %18
where $b_m\ne0$ is a contant and $p_m$ is a natural number.
Eq.(18) is important for the development of the Levinson theorem
in the next section.

For comparison we give the corresponding results in three dimensions.
The phase shifts are denoted by $\delta_l(k)$. By similar analysis
it can be shown in a potential $V_a(r)$ that
$$\tan\delta_l\to c_l\rho^{2q_l-1},\quad l=0,1,2,\ldots\quad
(k\to 0)\eqno(18')$$
where $c_l\ne0$ is a constant, and $q_l$ is a natural number for
$l\ne 0$, while $q_0$ may be a natural number or zero. We see that
$\delta_l(0)$ generally equals a multiple of $\pi$ for all $l$.
But $\delta_0(0)$ gets an additional $\pi/2$ when $q_0=0$. The latter
case does not occur for any $\eta_m(0)$, which is obvious from (18).
The difference between (18) and ($18'$) comes from the fact that the
Neumann function in the solution (14) involves the logarithmic
function while the spherical Neumann function in the three-dimensional
solution does not. It can be shown that $\delta_0(0)$ gets an
additional $\pi/2$ when there exists a half-bound state(a zero-energy
resonance) in the angular momentum channel $l=0$.

\section*{\large \S4. The Levinson theorem}

Now we proceed to establish the Levinson theorem by the Green
function method. Introduce the retarded Green function
$G({\bf r}, {\bf r'}, E)$ defined by
\begin{equation}
G({\bf r}, {\bf r'}, E)=\sum_\nu{\psi_\nu({\bf r})\psi^*_\nu
({\bf r'})\over E-E_\nu+i\epsilon} \end{equation}         %19
where $\{\psi_\nu({\bf r})\}$ is a complete set of orthonormal
solutions to (1), and $\epsilon=0^+$.
$G({\bf r}, {\bf r'}, E)$ satisfies the equation
\begin{equation}
(E-H+i\epsilon)G({\bf r}, {\bf r'}, E)=\delta({\bf r}-{\bf r'}).
\end{equation}         %20
For a free particle we have a similar definition:
\begin{equation}
G^{(0)}({\bf r}, {\bf r'}, E)=\sum_\nu{\psi^{(0)}_\nu({\bf r})
\psi^{(0)*}_\nu
({\bf r'})\over E-E^{(0)}_\nu+i\epsilon} \end{equation}         %21
where $\{\psi^{(0)}_\nu({\bf r})\}$
is a complete set of orthonormal solutions to Eq.(1) with $V=0$.
$G^{(0)}({\bf r}, {\bf r'}, E)$ satisfies
\begin{equation} (E-H_0+i\epsilon)G^{(0)}({\bf r}, {\bf r'}, E)
=\delta({\bf r}-{\bf r'})  \end{equation}         %22
where $H_0$ is the Hamiltonian of the free particle. We have the
integral equation for $G({\bf r}, {\bf r'}, E)$:
\begin{equation}
G({\bf r}, {\bf r'}, E)-G^{(0)}({\bf r}, {\bf r'}, E)=
\int d{\bf r}''\,G^{(0)}({\bf r}, {\bf r''}, E)V({\bf r''})
G({\bf r''}, {\bf r'}, E).\end{equation}            %23
In a central field $V({\bf r})=V(r)$(not necessarily $V_a(r)$),
we have
$$ \psi_\nu(r,\theta)=\psi_{m\kappa}(r,\theta)={u_{|m|\kappa}(r)
\over \sqrt r}{e^{im\theta}\over\sqrt{2\pi}},\quad m=0,\pm1,\pm2,
\ldots $$
where $\kappa$ is a quantum number associated with the energy
$E_{m\kappa}$ which is
determined by solving the radial equation
\begin{equation}
u''_{m\kappa}+\left[{2\mu\over\hbar^2}(E_{m\kappa}-V)-{m^2-1/4
\over r^2}
\right]u_{m\kappa}=0,\quad m=0,1,2,\ldots\end{equation}          %24
with appropriate boundary conditions in the radial direction. The
radial wave functions $u_{m\kappa}(r)$ satisfy the orthonormal
condition
\begin{equation}
(u_{m\kappa}, u_{m\kappa'})=\int_0^\infty dr\,u_{m\kappa}^*(r)
u_{m\kappa'}(r)=\delta_{\kappa\kappa'}.\end{equation}              %25
For an attractive field we have discrete spectrum($E_{m\kappa}<0$)
as well as continuous spectrum ($E_{m\kappa}>0$). We can, however,
require the wave functions to vanish at a sufficiently large radius
$R$($R\gg a$ for $V_a(r)$) and thus discretize the continuous part of
the spectrum.
In this case the upper limit of the integration in (25) should be
replaced by $R$. It is easy to show that
\begin{equation}
G({\bf r}, {\bf r'}, E)=G_0(r,r',E){1\over 2\pi}+\sum_{m=1}^\infty
G_m(r,r',E){\cos m(\theta-\theta')\over\pi}\end{equation}       %26
where
\begin{equation}
G_m(r,r',E)=\sum_{\kappa}{u_{m\kappa}(r)u_{m\kappa}^*(r')\over
\sqrt{rr'}(E-E_{m\kappa}+i\epsilon)},\quad m=0,1,2,\ldots.
\end{equation}                %27
For a free particle, the following results can be obtained in the same
way.
\begin{equation}
G^{(0)}({\bf r}, {\bf r'}, E)=G_0^{(0)}(r,r',E){1\over 2\pi}+
\sum_{m=1}^\infty G_m^{(0)}(r,r',E){\cos m(\theta-\theta')
\over\pi}\end{equation}       %28
where
\begin{equation}
G_m^{(0)}(r,r',E)=\sum_{\kappa}{u_{m\kappa}^{(0)}(r)u_{m\kappa}
^{(0)*}(r')\over
\sqrt{rr'}(E-E_{m\kappa}^{(0)}+i\epsilon)},\quad m=0,1,2,\ldots
\end{equation}                %29
where $u_{m\kappa}^{(0)}(r)$ satisfies (24) with $V=0$, and the energy
spectrum($E_{m\kappa}^{(0)}>0$) is discretized according to the above
described prescription. Thus the orthonormal relation for
$u_{m\kappa}^{(0)}(r)$ is similar to (25). Substituting (26) and (28)
into (23) we get an integral equation for $G_m(r,r',E)$:
\begin{equation}
G_m(r,r',E)-G_m^{(0)}(r,r',E)=\int dr''\,r''G_m^{(0)}(r,r'',E)
V(r'')G_m(r'',r',E),\quad m=0,1,2,\ldots.\end{equation}

Using the orthonormal relation (25) it is easy to show that
\begin{equation}
\int dr\,rG_m(r,r,E)=\sum_\kappa{1\over E-E_{m\kappa}+i\epsilon}.
\end{equation}               %31
Employing the mathematical formula
\begin{equation} {1\over x+i\epsilon}=P{1\over x}-i\pi\delta(x)
\end{equation}
and taking the imaginary part of the above equation we have
\begin{equation}
{\rm Im}\int dr\,rG_m(r,r,E)=-\pi\sum_\kappa\delta(E-E_{m\kappa}).
\end{equation}               %33
Integrating this equation over $E$ from $-\infty$ to $0^-$ yields
\begin{equation}
{\rm Im}\int_{-\infty}^{0^-}dE\,\int dr\,rG_m(r,r,E)=-n_m^-\pi
\end{equation}               %34
where $n_m^-$ is the number of bound states with negative energies and
with angular momentum $m\hbar$(when $m\ne0$ we have the same number of
bound states with angular momentum $-m\hbar$ as well). The possibility
of a zero-energy bound state will be discussed in the next section.
In a similar way one can show that
\begin{equation}
{\rm Im}\int_{-\infty}^{0^-}dE\,\int dr\,rG_m^{(0)}(r,r,E)=0.
\end{equation}               %35
Here and in (34) the integration over $E$ is performed to the upper
limit $0^-$ instead of 0 such that it suffers no ambiguity.
Combining (34) and (35) we have
\begin{equation} {\rm Im}\int_{-\infty}^{0^-}dE\,\int dr\,r[G_m(r,r,E)
-G_m^{(0)}(r,r,E)]=-n_m^-\pi.
\end{equation}               %36
On the other hand, substituting (27) and (29) into the rhs of (30)
we have
\begin{equation}
\int dr\,r[G_m(r,r,E)-G_m^{(0)}(r,r,E)]=\sum_{\kappa\sigma}
{(u_{m\sigma},u_{m\kappa}^{(0)})(u_{m\kappa}^{(0)}, Vu_{m\sigma})
\over (E-E_{m\kappa}^{(0)}+i\epsilon)(E-E_{m\sigma}+i\epsilon)}.
\end{equation}   %37
Using the radial equations for $u_{m\kappa}^{(0)}(r)$ and
$u_{m\sigma}(r)$ and the boundary condition that these radial wave
functions vanish at $r=0$ and $r=R$, it is not difficult to
show that
\begin{equation}
(u_{m\kappa}^{(0)}, Vu_{m\sigma})=(E_{m\sigma}-E_{m\kappa}^{(0)})
(u_{m\kappa}^{(0)}, u_{m\sigma}).\end{equation}                   %38
Substituting this result into (37) and taking the imaginary part we
have
\begin{equation}
{\rm Im}\int dr\,r[G_m(r,r,E)-G_m^{(0)}(r,r,E)]=\pi\sum_{\kappa
\sigma}[\delta(E-E_{m\kappa}^{(0)})-\delta(E-E_{m\sigma})]
|(u_{m\kappa}^{(0)}, u_{m\sigma})|^2.\end{equation}              %39
Integrating this equation over $E$ from $-\infty$ to $+\infty$
it is easy to find that
\begin{equation}
{\rm Im}\int_{-\infty}^{+\infty}dE\,\int dr\,r[G_m(r,r,E)
-G_m^{(0)}(r,r,E)]=0.\end{equation}                 %40
Similar to the three-dimensional case, this equation means that the
total number of states in a specific angular momentum channel is not
altered by an attractive field, except that some scattering states
are pulled down into the bound-state region. This result, together
with (36), leads to
\begin{equation} {\rm Im}\int_{0^-}^{+\infty}dE\,\int dr\,r[G_m(r,r,E)
-G_m^{(0)}(r,r,E)]=n_m^-\pi.\end{equation}                 %41
We have thereupon finished the first step in our establishment of
the Levinson theorem.

The next step  is to calculate the lhs of (41) in another way. In the
above treatment we have discretized the continuous spectrum of
$E_{m\kappa}^{(0)}$ and the continuous part of $E_{m\kappa}$. In the
following we will directly deal with these continuous spectra. We will
denote $u_{m\kappa}^{(0)}(r)$ by $u_{mk}^{(0)}(r)$,
and $u_{m\kappa}(r)$ with continuous $\kappa$ by $u_{mk}(r)$,
whereas those $u_{m\kappa}(r)$
with discrete $\kappa$(bound states) will be denoted by the original
notation.
The notations $E_{m\kappa}^{(0)}$ and $E_{m\kappa}$
with continuous $\kappa$ will also be changed to $E_{mk}^{(0)}$ and
$E_{mk}$(both are equal to $\hbar^2k^2/2\mu$) respectively.
The orthonormal relation for $u_{mk}^{(0)}(r)$ now takes
the form
\begin{equation} (u_{mk}^{(0)}, u_{mk'}^{(0)})=\delta(k-k').
\end{equation}
The orthonormal relation for $u_{mk}(r)$ is similar, while that for
$u_{m\kappa}(r)$ has the same appearance as (25). It is easy to show
that
$$ u_{mk}^{(0)}(r)=\sqrt{kr} J_m(kr),\quad k=\sqrt{2\mu E_{mk}^{(0)}}
/\hbar$$
satisfies the radial equation with $V=0$ and the orthonormal relation
(42). Thus $u_{mk}^{(0)}(r)$ has the asymptotic form
\begin{equation} u_{mk}^{(0)}(r)\stackrel{r\to\infty}{\longrightarrow}
\sqrt{2\over \pi} \cos\left(kr-{m\pi\over 2}-{\pi\over 4}\right)
\end{equation}
corresponding to (8). In an external field, the wave functions are
distorted and thus the asymptotic form for $u_{mk}(r)$ becomes
\begin{equation} u_{mk}(r)\stackrel{r\to\infty}{\longrightarrow}
\sqrt{2\over \pi} \cos\left[kr-{m\pi\over 2}-{\pi\over 4}
+\eta_m(k)\right] \end{equation}
corresponding to (9). Note that the coefficient in the asymptotic
form is the same for $u_{mk}^{(0)}(r)$ and $u_{mk}(r)$. In this
treatment it can be shown that
\begin{equation}
G_m(r,r',E)=\sum_{\kappa}{u_{m\kappa}(r)u_{m\kappa}^*(r')\over
\sqrt{rr'}(E-E_{m\kappa}+i\epsilon)}+
\int dk\,{u_{mk}(r)u_{mk}^*(r')\over
\sqrt{rr'}(E-E_{mk}+i\epsilon)}  \end{equation}              %45
and thus
\begin{equation}
{\rm Im}\int dr\,rG_m(r,r,E)=-\pi\sum_\kappa\delta(E-E_{m\kappa})
-\pi\int dk\,\delta(E-E_{mk})(u_{mk}, u_{mk}).  \end{equation}
Integrating this equation over $E$ from $0^-$ to $+\infty$ and taking
into account the fact that $E_{m\kappa}<0$ while $E_{mk}\ge0$, we have
\begin{equation} {\rm Im}\int_{0^-}^{+\infty}dE\,\int dr\,rG_m(r,r,E)=
-\pi\int dk\,(u_{mk}, u_{mk}).  \end{equation}      %47
There is no ambiguity in the integration over $E$ since the lower
limit is set to be $0^-$ instead of 0(note that $E_{mk}$ may equal 0).
As before, the possibility of a zero-energy bound state will be
discussed in \S5.  In the same way, we have
\begin{equation}
{\rm Im}\int_{0^-}^{+\infty}dE\,\int dr\,rG_m^{(0)}(r,r,E)=
-\pi\int dk\,(u_{mk}^{(0)}, u_{mk}^{(0)}).  \end{equation}      %48
It should be pointed out that the integrands in both (47) and (48)
are $\delta(0)(=\infty)$ according to Eq.(42) and a similar equation
for $u_{mk}(r)$. However, there is a subtle difference between
these two $\delta$ functions, and it is this difference that leads
to the Levinson
theorem. Since both integrands are singular, we first evaluate
\begin{equation}
(u_{mk}, u_{ml})_{r_0}-(u_{mk}^{(0)}, u_{ml}^{(0)})_{r_0}\equiv
\int_0^{r_0}dr\,u_{mk}^*(r)u_{ml}(r)-
\int_0^{r_0}dr\,u_{mk}^{(0)*}(r)u_{ml}^{(0)}(r)\end{equation}
where $r_0$ is a large but finite radius, and finally take the limit
$l\to k$ and $r_0\to\infty$.  As in the discretized case, we have the
boundary condition
\begin{equation} u_{mk}^{(0)}(0)=0, \quad u_{mk}(0)=0.
\end{equation}      %50
Using the radial equation and this boundary condition it is easy
to show that
\begin{equation}
(k^2-l^2)(u_{mk}, u_{ml})_{r_0}=u_{mk}^*(r_0)u'_{ml}(r_0)
-u_{mk}^{\prime*}(r_0)u_{ml}(r_0).\end{equation}
Since $r_0$ is large, we can use (44) to evaluate the rhs and in the
limit $l\to k$ we get
\begin{equation}
(u_{mk}, u_{mk})_{r_0}={r_0\over \pi}+{1 \over \pi}\eta'_m(k)-
{(-)^m\over 2\pi k}\cos[2kr_0+2\eta_m(k)].\end{equation}
In the same way we have
\begin{equation} (u_{mk}^{(0)}, u_{mk}^{(0)})_{r_0}={r_0\over \pi}-
{(-)^m\over 2\pi k}\cos2kr_0.\end{equation}
Therefore,
\begin{equation}
(u_{mk}, u_{mk})_{r_0}-(u_{mk}^{(0)}, u_{mk}^{(0)})_{r_0}=
{1\over\pi}\eta'_m(k)+{(-)^m\over 2}\delta(k)\sin2\eta_m(k)+
{(-)^m\over\pi k}\cos 2kr_0\sin^2\eta_m(k)\end{equation}    %54
where we have employed the well-known formula
$$ \lim_{r_0\to\infty}{\sin 2kr_0\over \pi k}=\delta(k). $$
So far in this section $V(r)$ need not be $V_a(r)$. In the following
we set $V(r)=V_a(r)$. Then Eq.(18) is valid, and we have
$\sin2\eta_m(0)=0$. Therefore the second
term on the rhs of (54) vanishes. Integrating this result over $k$
(from 0 to $+\infty$), taking the limit $r_0\to \infty$, and
incorporating the results (47), (48) we arrive at
\begin{eqnarray}
&&{\rm Im}\int_{0^-}^{+\infty}dE\,\int dr\,r[G_m(r,r,E)
-G_m^{(0)}(r,r,E)]  \nonumber\\
&&=\eta_m(0)-\eta_m(\infty)-(-)^m\lim_
{r_0\to\infty}\int_0^\infty dk\,{\cos 2kr_0\over k}\sin^2\eta_m(k).
\end{eqnarray}       %55
The last term in this equation can be decomposed into two integrals,
the first from 0 to $\varepsilon=0^+$, which can be shown to vanish
on account of (18), while the second from $\varepsilon$ to $+\infty$,
which also vanishes in the limit $r_0\to\infty$ since the factor
$\cos2kr_0$ oscillates very rapidly. Therefore we have
\begin{equation} {\rm Im}\int_{0^-}^{+\infty}dE\,\int dr\,r[G_m(r,r,E)
-G_m^{(0)}(r,r,E)]=\eta_m(0)-\eta_m(\infty).\end{equation}        %56
Combining (56) and (41) we arrive at the Levinson theorem:
\begin{equation}\eta_m(0)-\eta_m(\infty)=n_m^-\pi,\quad m=0,1,2,
\ldots. \end{equation}
In the next section we will discuss some aspects of this theorem,
and get a more general form which takes zero-energy bound states
into consideration.

\section*{\large \S5. Discussions}

In this section we discuss and clarify some aspects of the Levinson
theorem obtained in the form (57) in the last section.

1. {\it On zero-energy bound states}. In \S4 we have not taken
into account
the possible existence of a zero-energy bound state. Indeed, this may
occur for a square well potential with radius $a$ and depth $V_0$ when
$m>1$ and $J_{m-1}(\xi)=0$ where $\xi=k_0a$ and $k_0=\sqrt{2\mu V_0}/
\hbar$. (For $m=0,1$ regular solutions with zero energy may be found
but they are not normalizable and thus are not bound states.
see below) The
existence of a zero-energy bound state would not alter the results
(34)-(36) where $n_m^-$ is the number of bound states with negative
energies. Therefore Eq.(41) remains valid in this case. On the other
hand, Eq.(47) becomes
\begin{equation} {\rm Im}\int_{0^-}^{+\infty}dE\,\int dr\,rG_m(r,r,E)=
-\pi-\pi\int dk\,(u_{mk}, u_{mk})  \end{equation}      %58
which is an obvious consequence of (46). Accordingly, the result
(56) becomes
\begin{equation} {\rm Im}\int_{0^-}^{+\infty}dE\,\int dr\,r[G_m(r,r,E)
-G_m^{(0)}(r,r,E)]=\eta_m(0)-\eta_m(\infty)-\pi.\end{equation}    %59
Hence, the Levinson theorem takes in the present case the form
\begin{equation}\eta_m(0)-\eta_m(\infty)=n_m\pi   \end{equation}
where $n_m=n_m^-+1$ represents the total number of bound
states, including the one with zero energy. When there is no
zero-energy bound state, $n_m=n_m^-$.
Thus Eq.(60) holds in any case and is the final form of the Levinson
theorem.

To see the difference in zero-energy states between two and three
dimensions, we notice that the two-dimensional radial wave function
$u_{m\kappa}(r)$ satisfies (24). When regarded as a one-dimensional
Schr\"odinger equation, the effective potential reads
$$\tilde V_2(r)=V(r)+{\hbar^2(m^2-1/4)\over 2\mu r^2}$$
where the subscript ``2'' indicates two dimensions. In three
dimensions, with $\psi_{lm\kappa}(r,\theta,\phi)=r^{-1}\chi_{l\kappa}
(r)Y_{lm}(\theta,\phi)$(here $r$, $\theta$ etc. should not be
confused with those in two dimensions), the radial wave function
$\chi_{l\kappa}(r)$ satisfies
$$\chi_{l\kappa}''+\left[{2\mu\over\hbar^2}(E_{l\kappa}-V)-{l(l+1)
\over r^2}\right]\chi_{l\kappa}=0,\quad l=0,1,2,\ldots.
\eqno(24')$$
When this is regarded as a one-dimensional Schr\"odinger equation,
the effective potential is
$$\tilde V_3(r)=V(r)+{\hbar^2 l(l+1)\over 2\mu r^2}$$
which is obviously different from $\tilde V_2(r)$, and as a
consequence
the zero-energy solutions are different from those of (24). More
specifically, with $V(r)=V_a(r)$ and $E_{m0}=0$, the exterior
solution($r>a$) of (24) reads
$$u_{m0}^+(r)=r^{-m+1/2}.$$
A zero-energy solution exists if $V_a(r)$ is such that the interior
solution $u_{m0}^-(r)$ can be connected with $u_{m0}^+(r)$ at $r=a$
smoothly. This leads to the above mentioned condition for a square
well potential. Here we are concerned about the normalizability of
the solution. It is clear that the above solution can be normalized
and thus is a  bound state only when $m>1$. We have
$\psi_{10}\stackrel{r\to\infty}{\longrightarrow}0$, and $\psi_{00}$
is finite at infinity though
$u_{00}(r)\stackrel{r\to\infty}{\longrightarrow}\infty$, thus both
are regular. But either of them can not be normalized and thus is not
a bound state. In this case $\eta_m(0)$ gets no additional term and
Eq.(57) is not modified. When $m>1$, however, $n_m^-\to n_m^-+1=n_m$
and $\eta_m(0)$ gets an additional $\pi$ if a zero-energy solution%
(a bound state) actually exists. In three dimensions,
with $V(r)=V_a(r)$ and $E_{l0}=0$, the exterior
solution of ($24'$) reads
$$\chi_{l0}^+(r)=r^{-l}.$$
This is normalizable when $l>0$. Thus only $l=0$ is distinguished.
When the $l=0$ solution really emerges, $\delta_0(0)$ gets an
additional $\pi/2$ as pointed out in \S3, and the Levinson theorem
is modified by $n_0\to n_0+1/2$. This is different from the case
$m=0,1$ in two dimensions where $\eta_m(0)$ always equals a multiple
of $\pi$. When $l>0$ the case is basically the same as $m>1$
in two dimensions.
                                    
2. {\it About $\eta_m(\infty)$}.  Write down two radial equations
for $u_{mk}(r)$ and $\tilde u_{mk}(r)$ in the external potentials
$U(r)$ and $\tilde U(r)$ respectively. Using the boundary condition
(50) and the asymptotic form (44) for $u_{mk}(r)$ and similar ones
for $\tilde u_{mk}(r)$, it is easy to show that
\begin{equation}
\sin[\eta_m(k)-\tilde \eta_m(k)]=-{\pi\mu\over\hbar^2k}
\int_0^\infty dr\,[U(r)-\tilde U(r)] u_{mk}^*(r)\tilde u_{mk}(r).
\end{equation}
Now we take $U(r)=V(r)$(not necessarily be $V_a(r)$)
and $\tilde U(r)=0$, it is natural to define
$\tilde\eta_m(k)=0$ in the absence of an external field. Obviously,
$\tilde u_{mk}(r)=u_{mk}^{(0)}(r)$, so (61) becomes
\begin{equation} \sin\eta_m(k)=-{\pi\mu\over\hbar^2k}
\int_0^\infty dr\,u_{mk}^*(r)V(r)u_{mk}^{(0)}(r).\end{equation}   %62
For very large $k$, $V(r)$ can be ignored in the radial equation since
it is less singular than $r^{-2}$ near the origin and well behaved
elsewhere as assumed and so can be neglected everywhere(cf. Eq.(24)).
Therefore $u_{mk}(r)$ in (62) can be replaced by $u_{mk}^{(0)}(r)$
in this limit, and we have for very large $k$
\begin{equation} \sin\eta_m(k)=-{\pi\mu\over\hbar^2k}
\int_0^\infty dr\,|u_{mk}^{(0)}(r)|^2V(r).\end{equation}      %63
Substituting $u_{mk}^{(0)}(r)=\sqrt{kr}J_m(kr)$ into this equation,
and using the asymptotic formula for the Bessel functions at large
argument,  we have approximately
\begin{equation} \sin\eta_m(k)=-{2\mu\over\hbar^2k}
\int_0^\infty dr\,\cos^2\left(kr-{m\pi\over2}-{\pi\over4}\right)V(r)
.\end{equation}      %64
Further, since $k$ is very large, the cosine oscillates very rapidly
and thus the squared cosine may be replaced by its mean value 1/2.
In this way we arrive at
\begin{equation} \sin\eta_m(k)=-{\mu\over\hbar^2k}
\int_0^\infty dr\,V(r)\end{equation}      %65
for very large $k$, where we assume that the integral exists. This
result has the same form as that in three dimensions.
Of course it holds in the special case $V(r)=V_a(r)$. Obviously,
$\sin\eta_m(k)\to0$ when $k\to\infty$, hence we can freely define
$\eta_m(\infty)=0$. Under this convention the Levinson theorem
takes the form
\begin{equation} \eta_m(0)=n_m\pi,\quad m=0,1,2,\ldots. \end{equation}
This means that the phase shift at threshold serves as a counter for
the bound states. This is similar to the case in three dimensions, but
is somewhat simpler. In three dimensions, the case $l=0$ should be
modified when there exists a zero-energy resonance(a
half-bound state). Here we have no such problem.

3. {\it Extension to more general potentials}. In the above
development
of the Levinson theorem, we have assumed that $V(r)=0$ when $r>a$.
We also assumed that $V(r)$ is less singular than $r^{-2}$ when
$r\to 0$, so that the power dependence in (13) near $r=0$ is valid.
Indeed, the existence of the integral in (65) requires that $V(r)$
is less singular than $r^{-1}$ when $r\to 0$. We will always assume
that $V(r)$ is sufficiently well behaved near the origin such that
the above requirements are all satisfied. On the other hand,
the radius
$a$ beyond which $V(r)$ vanishes is not specified in our discussion.
Though both $\eta_m(0)$ and $n_m$ in (66) depend on the particular
form of $V(r)$ and thus depend on $a$, the equality between them does
not. Hence one expects that (66) remains valid when $V_a(r)$ is varied
continuously to the limit $a\to\infty$ if $n_m$ remains finite in the
process. It seems that the Levinson theorem holds for quite general
potentials(well behaved near the origin as emphasized above) as long
as they decrease rapidly enough when $r\to \infty$ such that the total
number of bound states in a particular angular momentum channel
is finite.

Extension of the present work to relativistic quantum mechanics is
currently under progress.

\vskip 2pc

The author is grateful to Prof. Guang-jiong Ni for useful
communications and for encouragement.
This work was supported by the
National Natural Science Foundation of China.

\newpage
\parindent 0pc
{\large References}\newline
[1] N. Levinson, K. Dan. Vidensk. Selsk. Mat.-Fys. Medd. 25,
No.9(1949).

[2] J. M. Jauch, Helv. Phys. Acta. 30, 143(1957).

[3] A. Martin, Nuovo Cimento 7, 607(1958).

[4] R. G. Newton, J. Math. Phys. 18, 1348; 1582(1977); \newline
{\it Scattering Theory of Waves and Particles}(McGraw-Hill, New York,
1966).

[5] G.-J. Ni, Phys. Energ. Fort. Phys. Nucl. 3, 432(1979).

[6] Z.-Q. Ma and G.-J. Ni, Phys. Rev. D31, 1482(1985).

[7] N. Poliatzky, Phys. Rev. Lett. 70, 2507(1993);
Helv. Phys. Acta. 66, 241(1993).

[8] J. Piekarewicz, Phys. Rev. C48, 2174(1993).
\end{document}